\documentclass[11pt]{article}
\usepackage{epsfig, graphicx, amsmath, amssymb}
\begin{document}
\title{Gravitational wave constraints on corrections to Bekenstein-Hawking Area Formula in classical F(R) gravity and quantum GR : implications for theory parameters.}
\author{Parthasarathi Majumdar\footnote{bhpartha@gmail.com} \\School of Physical Sciences \\Indian Association for the Cultivation of Science, Kolkata 700032, India}
\maketitle
\begin{abstract}

This contribution considers constraints from analyses of gravitational wave data from binary black hole coalescence, on possible corrections to the Bekenstein-Hawking Area Formula for black hole entropy. Most recent analyses of gravitational wave data from the LVK Consortium appear to confirm the Hawking Area Theorem for black holes at a $5 \sigma$ accuracy, for the `loudest' signal (SNR of the order of $80$) of binary black hole merger inherent in the recent observation GW250114. Amalgamating this result with Bekenstein's ideas of black hole entropy and the generalized second law of thermodynamics, we constrain leading inverse area corrections for large horizon area spherical black hole solutions of classical $F(R)$ gravity, using the Wald entropy function formalism. The implementation of the observational constraints entails the notion of `absolute consistency' which we introduce and contrast with `relative consistency'. This absolute consistency criterion is shown to relate some of the parameters of $F(R)$ gravity. Next we consider  leading quantum general relativistic corrections to the Area formula, arising both from the non-perturbative matter-free Loop Quantum Gravity and matter-dependent, perturbative  Entanglement entropy approaches. Combining the leading logarithmic corrections in horizon area (for large areas) from both approaches, and imposing absolute consistency with the observational validation of the Area Theorem, is shown to lead to significant restrictions on the spin and number of species of Beyond Standard Model spectrum of elementary particles, some of which are often assumed to be Dark Matter candidates.    
\end{abstract} 
%\end{document}
\section{Introduction}

Why at all should gravitational wave (GW) data from binary black hole coalescence (BBHC) constrain corrections to the Bekenstein-Hawking Area Formula (BHAF) for black hole entropy, computed theoretically from classical modified gravity theories or quantum general relativity proposals ? The vast difference in energy scales of the observed merger events and that of the theoretical computation of corrections to the BHAF, should apparently rule out such a possibility. The reason that this is not the case has to do with the incredible accuracy ($5 \sigma$) with which the general relativistic  Hawking Area Theorem (HAT) for BBHCs has been validated from GW data \cite{teu20}-\cite{bad22}. For astrophysical black holes of large horizon area, it is perhaps reasonable that corrections to the BHAF, small as they ought to be within any theoretical framework, may be construed to be {\it additive} \cite{pm2024}-\cite{pm2026}. This assumption, and Bekenstein's (and Hawking's) \cite{bek73}-\cite{haw75} ideas on black hole entropy and the generalized second law of thermodynamics, permit us to {\it bound} the additive corrections on the basis of GW observational results. The bound obtained in this manner is nontrivial only if an {\it absolute consistency criterion} is adopted, whereby the algebraic sign of the corrections (rather than their magnitude, which is small anyway) is given primacy. 

The implications of such a criterion are rather unexpected, especially on the parameters of the theory used to compute the corrections. Certain issues pertaining to the Large Hadron Collider (LHC) perhaps may demystify this to some extent. The remarkable confirmation of the Standard Strong-Electroweak theory (SSET) of fundamental particle interactions is achieved at the LHC by first ascertaining the twenty five odd free parameters of the theory, while ignoring completely the gravitational interactions. However, from a cosmological standpoint, the events created at the LHC might be taken to correspond to an earlier epoch in the evolution of the universe when it was much smaller in size than today's universe. With the bigger energies at the disposal of particles in that epoch, is it still correct to ignore all effects of (quantum) gravity. Standard wisdom, based on perturbative arguments would tend to reinforce that already gravity has ceased to be of importance in such an epoch. On the other hand, the large number of free parameters in the SSET corresponds to the {\it unknown} aspects of the theory, where phenomena like parity violation, CP violation, baryon asymmetry, fermion mass spectrum etc exist with no theoretical understanding until today. We have done little more than merely parametrise these features of the SSET. Perhaps quantum gravity may play a role in unravelling this unknown sector of the SSET. It is of importance to explore if this is indeed the case. 

This paper considers two disparate examples to illustrate the basic tenet, namely classical black holes in modified $F(R)$ gravity, and quantum black holes in both perturbative and non-perturbative quantum general relativity. 

In the case of modified $F(R)$ gravity theories for a large class of smooth functions of the Ricci scalar, assumed to be regular at the origin of the argument, which admit spherical black hole solutions of the modified Einstein equations, corrections to the BHAF are computed \cite{dmm2026} via the Wald-Jacobson-Kang-Myers entropy function formalism \cite{rmw1992}-\cite{jkm1994}. GW observational results via the absolute consistency criterion of inverse power (in horizon area) corrections to the BHAF, then lead to constraints on various derivatives of $F(R)$ for small arguments appropriate to astrophysical black holes. These constitute a class of parameters of the theory, which are a priori independent; the absolute consistency criterion on the inverse power law corrections relates some of them, unexpectedly. 

Likewise, for quantum black holes in quantum general relativity, both non-perturbative, background-independent quantum fluctuations of matter-free spacetime, and background-dependent, perturbative quantum fluctuations of matter fields and spacetime are concomitantly considered. For the background-independent, non-perturbative avenue we follow the approach of Loop Quantum Gravity (LQG) \cite{km98}-\cite{abhi-pm14} based on the canonical quantization of {\it isolated horizons} using the $SU(2)$ Chern-Simons theory within the spin network basis. In the perturbative approach we follow the Callan-Wilczek-Solodukhin \cite{cal1994}-\cite{sol2020} approach to Entanglement Entropy (ENT) of matter field fluctuations in a classical black hole background. One loop results in this approach include the leading logarithmic corrections to the BHAF upon suitable normalization of the Newton gravitational constant. For macroscopically large (astrophysical) black holes, leading logarithmic corrections to BHAF substantively restrict the number of species for each spin of matter field fluctuations. 

This review is organized as follows : in section 2, we discuss black hole horizons and focus on the HAT enunciation. We also note the implication of the HAT for BBHCs into a remnant black hole. We next 
summarize the results of the latest analyses of GW data corresponding to the very recently observed event GW250114, focusing on the validation of the HAT for BBHCs leaving a similar remnant black hole. We do not delve into extreme mass ratio inspirals. The next section (3) motivates the notion of black hole entropy from the statistical mechanics of isolated systems, taken together with basic tenets of vacuum GR. The BHAF is motivated and illustrated for BBHCs. The Generalised Second Law (GSL) of Thermodynamics is clarified and distinguished from the Laws of Black Hole Mechanics, to obviate common misconceptions in the literature. We then briefly discuss possible sources of modification or corrections to the BHAF, when one moves away from classical GR, either remaining within classical spacetime geometry, or venturing into the realm of quantum GR proposals. The observational HAT validation is then amalgamated with the GSL, using the possibility of {\it additive} modifications to the BHAF. This amalgamation leads to restrictions on the corrections to the BHAF. Assuming that the magnitude of these corrections are always tiny, we focus on what may be an {\it absolute consistency criterion} with the observations. We argue that the alternative of a {\it relative} consistency criterion may not predict much in this situation, and is almost always satisfied. On the other hand, if the phenomenological implications of the absolute consistency criterion do not show up in the real world, our contentions are immediately made false. This is perhaps why this criterion has strength. The fourth section presents the basic tenets of the Wald-Iyer-Jacobson-Kang-Myers (WIJKM) formalism for the entropy of function of a black hole solution of any general coordinate invariant theory of gravity beyond GR. We then specialize to the case of $F(R)$ gravity \cite{bre2004}-\cite{mod2011}, without restricting the nature of the function of the Ricci scalar. However, we do restrict to static spherical vacuum black hole solutions of the theory,  with large (macroscopic) horizon areas relevant for astrophysical observations, which confines our attention to a large but not all-inclusive class of functions $F$.  We examine the predictions of the WIJKM formalism and extract from this the appropriate BHAF and possible inverse horizon area power law corrections, within the classical geometry framework. Next, the corrections are subjected to the absolute consistency criterion vis-a-vis the observational validation of the HAT. This leads to certain inequalities between various derivatives of $F(R)$ for vanishingly small Ricci scalar corresponding to large horizon areas. Section 5 considers leading logarithmic corrections to the BHAF within a {\it quantum gravity} scenario, for macroscopic black holes of large horizon area. We attempt to combine the non-perturbative LQG approach and the perturbative ENT approach for the coefficient of the logarithmic correction to the BHF. The computational result is then subjected to the absolute consistency criterion. The unexpected implications for the spectrum of particles beyond the SM are then discussed, especially in view of the fact that some of these particles are under consideration in the particle cosmology lierature as Dark Matter candidates. We summarise our conclusions and outlook in section 6. 

\section{Hawking Area Theorem}

\subsection{Theoretical Aspects}

\hglue 3cm
\includegraphics[height=6cm]{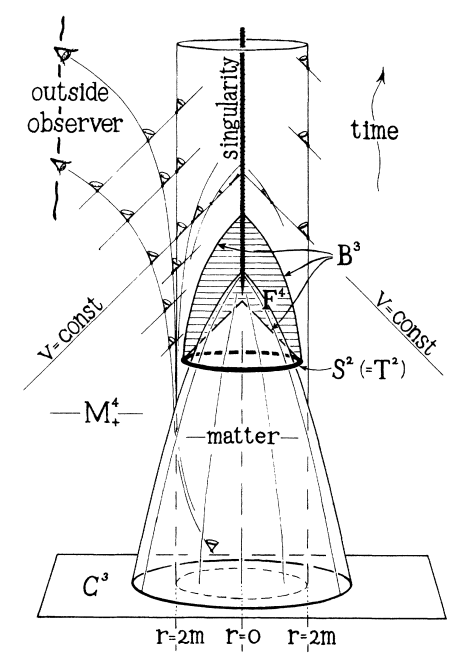}
\vglue -6cm
\hglue 8cm
\includegraphics[height=6cm]{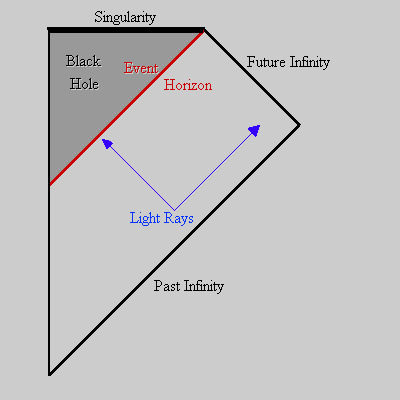}

\noindent {\small Fig. 1. The left panel depicts a spherical black hole in ingoing null coordinate frame \cite{pen65}, with trapped surface $S^2$, the event horizon and singularity. The right panel shows the conformal diagram for the same black hole}   
\vglue .1cm 

An asymptotically flat black hole spacetime is defined as
\begin{eqnarray}
{\cal B} &=& {\rm sptm}  - {\rm past}~({\rm future~null~infinity}~ {\cal I}^+) \nonumber \\
\partial {\cal B} &=& h~,~h \cap \Sigma_{1,2} \sim S^2 
\end{eqnarray} 
In the above equations, the boundary $h$ of the black hole spacetime is called the {\it event horizon}. For the left panel diagram, the vertical dotted line represents this horizon, especially after the time of formation of the trapped surface $S^2$. Imagine horizontal lines at each time interval of the time axis; each of these lines represent a spacelike hypersurface $\Sigma$ which cuts (`foliates') the black hole at a fixed time slice. Thus, $\Sigma_{1,2}$ represents two such time slices. It is obvious that if this represents an {\it isolated} black hole, then the area of the sphere $S^2$ obtained by any such foliation, is the same for any time slice. In other words, the area of the two-surface formed by foliation of the horizon by a spacelike hypersurface is a {\it conserved} quantity for an isolated black hole. 

On the other hand, if the black hole is {\it accreting} matter from its environment, like from a companion giant star, then the area of the 2-sphere for different foliations will no longer be conserved.     
\begin{center}
\includegraphics[height=7cm]{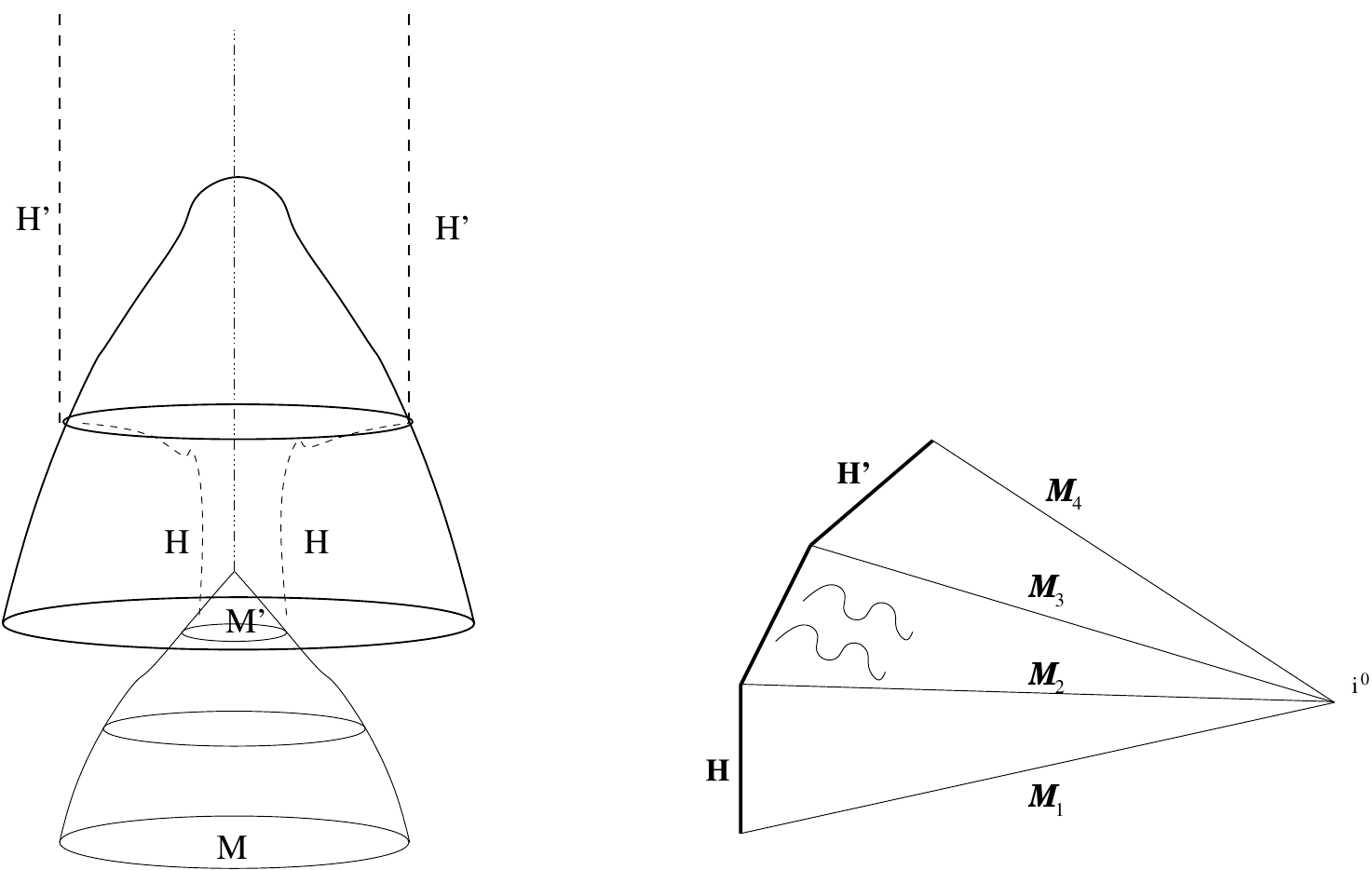}

{\small Fig. 2 Views of an accreting black hole from the ingoing null coordinate frame and the conformal frame. $M_{1,2,3,4}$ correspond to spacelike foliations $M_I$ indicated by $\Sigma$ above. }
\end{center}

Because of accreting matter from the neighbourhood, the area of the horizon changes : it invariably {\it increases} : $A_{fin} > A_{ini} \rightarrow \Delta A > 0$. This is the Hawking Area Theorem \cite{hat71}, which can be proven in general for all accreting black holes in general relativity. An immediate implication of this theorem is that a black hole can never decompose into a pair of black holes, while two black holes can certainly coalesce into a single remnant black hole. In the latter situation, i.e., for a BBHC, 
\begin{equation}
A_{h,rem} > A_{h,1} + A_{h,2} \Rightarrow (\Delta A_h/A_{h,ins}) > 0
\end{equation}  
We may mention that several proofs of the HAT are now available in graduate textbooks, which do not make use of the Einstein equation, but properties of geodesics and energy conditions. Thus, the domain of applicability of the HAT appropriate to BBDC GW data may be larger than strictly GR-based approaches. This gives us the rationale to use the very accurate results of recent observation of a BBHC event with an SNR close to $80$ (GW250114), to determine constraints on a class of modified gravity theories in the sequel. 

\subsection{GW250114 : the `loudest' GW signal with SNR $\sim 80$ \cite{abac2025}}

\begin{center}
\includegraphics[height=10cm]{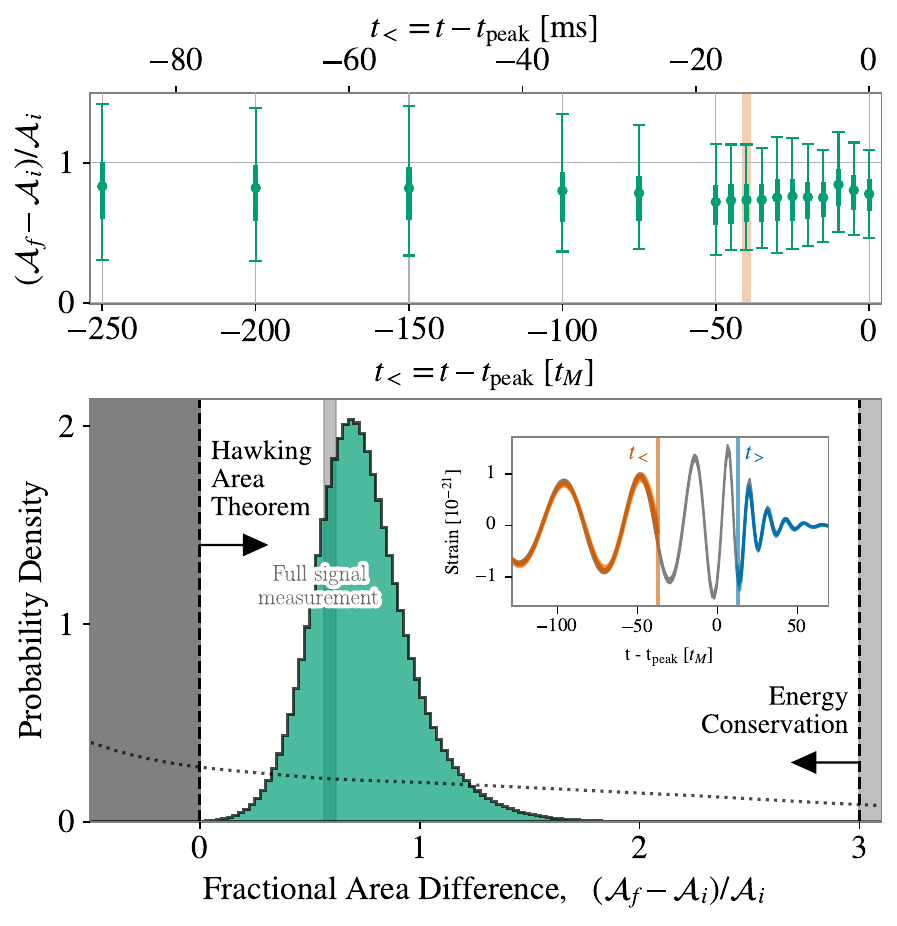}

{\small Fig.3 Top panel records fractional area increase in BBHC, as a function of inspiral time upto merger. Bottom panel records probabiliy distribution of fractional area increase in BBHC event GW250114} 
\end{center}

Earlier inferences on GW observational validation of the HAT, performed in ref.s \cite{teu20}-\cite{bad22} based mainly on GW150914, are thus strengthened manifolds through this latest observation and analysis in ref. \cite{abac2025}. The use of surrogates of numerical relativity techniques, now being performed widely, also testifies to the improvement made with respect to the days of the PPN techniques which worked very well for the inspiral phase of BBHCs, but proved inadequate for the merger and ringdown phases which required different techniques of analysis. 

The statistical significance of the validity of the HAT, as reported in ref. \cite{abac2025} for a Kerr-type black hole remnant, is $4.5$ standard deviations for  the overall posterior data, with the significance exceeding $5 \sigma$ for certain truncation choices in the inspiral-merger-ringdown time frame of the data. This significance is far greater than the earlier assays \cite{teu20}-\cite{bad22} which were closer to $2.5 \sigma$ at most. With this kind of statistical precision, the issue is whether constraints might be gleaned from the HAT-validating analysis of GW250114 data, on parameters of modified gravity  theories, focusing on calculated quantities of those theories which are {\it small} deviations from the general relativity result. The reason for the latter restricted choice is that none of the modified theories have been fully employed yet to analyze the GW250114 data, so that a complete consistency check of the data with any of those theories is still unavailable.

\section{Black Hole Entropy and the BHAF}

\subsection{Basics}

In ordinary thermodynamics : energy ${\cal E}$ is a conserved quantity for every isolated system; in equilibrium statistical mechanics this is characterized by a {\it microcanonical} ensemble. Within such an ensemble, In ordinary thermodynamics the energy of an isolated system ${\cal E}$ is a {\it conserved} quantity. In equilibrium statistical mechanics, such systems are usually described by a {\it microcanonical} ensemble. For such an ensemble, the entropy $S=S({\cal E}) > 0$ is defined as a positive function of the energy, with additive properties : $S({\cal E}_1 + {\cal E}_2) = S_1({\cal E}_1) + S_2({\cal E}_2)$ for two isolated systems with energies $E_1$ and $E_2$. The temperature of such a system is a derived quantity, and given by $T^{-1} = (\partial S/\partial E)$. When interactions are turned on {\it adiabatically}, i.e., however violent they may be, the final system is once again an isolated system in equilibrium, it is known that, if the energy changes ${\cal E}_{ini} \rightarrow {\cal E}_{fin} \Rightarrow S({\cal E}_{fin}) > S({\cal E}_{ini})$. This is basically the second law of thermodynamics.  

However, this changes for a universe with black holes present. If we define black holes as exact solutions of general relativistic vacuum Einstein's equations with certain isometries, then, such spacetimes cannot possibly have a bulk energy or momentum density at every point in them. The reason is simply that one can always transform to a local Lorentz frame where the metric is locally flat, and therefore cannot possibly support a local {\it pure gravitational} energy-momentum tensor. Another way of saying the same thing is that, for pure source-free general relativity, the energy and momentum densities vanish as first class constraints. Thus, the energy (or momentum) of a black hole has to be defined {\it globally}, as the Komar, or ADM or Bondi, mass (and  spin for spinning spacetimes). Thus, an isolated black hole cannot carry energy as such. However, as we have seen earlier, one can associate with an isolated black hole its horizon area (upon foliation by a spacelike hypersurface : a time slice) as a conserved quantity in the sense that the cross-sectional area has the same value on every time slice : $S_{bh} \equiv S_{bh}(A_h)$. The black hole entropy function thus defined has all the additive properties mentioned earlier, so long as isolated black holes are believed to be of constant horizon cross-sectional area. 

Now if the black hole is in an environment where it can accrete matter from a neighbouring star, then once accretion stops, the HAT would imply $A_{h,fin} > A_{h,ini}$. Bekenstein \cite{bek73} postulated that under these situations,
\begin{eqnarray}
S_{bh,ini} + S_{ext,ini} < S_{bh,fin} +S_{ext,fin} \label{gsl}
\end{eqnarray}
In other words, for a universe with black holes, the second law of thermodynamics should undergo a generalization, leading to what is known as the Generalized Second Law (GSL) of thermodynamics : the total entropy of black hole plus external matter can never decrease in any physical process (like accretion). If the matter entropy decreases because of accretion, this must be compensated by an increase in black hole entropy. 

For the BBHC events reported by LVK, there is no data on accretion of the inspiral or remnant black holes. In this situaton we assume that there is {\it no} accretion, so that eqn. (\ref{gsl}) is replaced by  
\begin{eqnarray}
S_{bh,1}(A_{h1}) + S_{bh,2}(A_{h2}) < S_{bh,rem}(A_{hr}) +S_{GW} \label{gslbb}
\end{eqnarray}
Here, $S_{GW}$ is the entropy of the GWs emitted during merger. On the basis of the data of GW150914, an estimate has been made of $S_{GW}$ in ref. \cite{mr21}, and we find that $S_{GW} << S_{bh}$. However, a better estimate than ours is certainly in order. For the rest of this review, we shall discard $S_{GW}$. 

Bekenstein \cite{bek73} made the further stipulation that black hole entropy must {\it scale} like the cross-sectional horizon area, so that in units where the Boltzmann constant can be set to unity,  
\begin{eqnarray}
S_{bh} = \frac{A_h}{A_{Pl}} \equiv S_{BH} \label{bhaf} 
\end{eqnarray} 
where $A_{Pl} = G \hbar/c^3$ is the Planck area (of order of magnitude $10^{-66}$ cm$^2$). Eqn.(\ref{bhaf}) is called the Bekenstein-Hawking Area Formula for black hole entropy, proposed first in ref. \cite{bek73}, and established as a consistency check with the Hawking radiation formula by Hawking \cite{haw75}. We have absorbed a factor of $4$ in a redefinition of $G$. The BHAF has been often confused with the HAT in various parts of the literature. Observe that while the HAT is a precise mathematical theorem within spacetime geometry with a rigorous proof, the BHAF is more or less a {\it postulate} whose precise verification depends on the {\it microstates} employed to define entropy. Bekenstein himself felt \cite{bek73} that an ab initio derivation may need states of a {\it quantum} theory of gravity as the microstates for this definition, because black holes in general relativity are often unique solutions of the vacuum Einstein equation, and as such cannot carry any entropy. However, considering serious proposals for quantum gravity, it stands to reason to expect {\it corrections} to the BHAF for black hole entropy, ensuing from the quantization scheme. For astrophysically large-horizon black holes of the type observed in LVK, such corrections may have a leading logarithmic (in horizon area) part, followed by subleading corrections having inverse powers in horizon area. 

Another manner of appearance of corrections to the BHAF may emerge from applying the WIJKM (see above) definition of an entropy function in terms of an integral of the classical  N\"other charge over a suitable codimension two cross-sectional foliation of the horizon of a black hole spacetime, in a modified classical theory of gravity. Here, the leading corrections are expected to be inverse powers of area terms, for large horizon areas. Thus, with both alternatives, we may have
\begin{eqnarray}
S_{bh} = S_{BH}(A_h) + s_{bh}(A_h,{\bf Q})~,~|s_{bh}| << S_{BH} \label{cor}
\end{eqnarray}    
where $s_{bh}$ represent corrections to the BHAF, with ${\bf Q}$ being additional charges that may arise in the changed mathematical structure  of GR, either due to quantization or due to classical modification. The aim in what follows is to examine if the theoretically computed corrections $s_{bh}$ can be constrained by GW observations validating the HAT. It is obvious that the validation of the HAT is tantamount to 
\begin{eqnarray}
S_{BH1} + S_{BH2} < S_{BHr} \Rightarrow \Delta S_{BH} \equiv  S_{BHr}-(S_{BH1}+S_{BH2}) > 0~.~ \label{HATN}
\end{eqnarray}
As we discuss in the next section, a validation of this relation to an accuracy of $5\sigma$ through observations must bound calculated $s_{bh}$ for BBHCs, once amalgamated with GSL.  

\subsection{GW Data and Absolute Consistency Criterion}

What is the nature of the entropy corrections ? We restrict our focus on astrophysical black holes of macroscopically large horizon cross-sectional areas. The correction $s_{bh}(A_h, {\bf Q})$ then can be expanded as a power series in inverse powers of the horizon area which is a small parameter. However, in certain quantum gravity proposals, in addition to power law corrections, the leading correction is always one which is logarithmic in the horizon area, leading to 
\begin{eqnarray}
s_{bh}(A_h, {\bf Q}) = s_0({\bf Q}) ~\log S_{BH} + s_1({\bf Q}) ~S_{BH}^{-1} + \cdots \label{corr}
\end{eqnarray}
where, the dots refer to higher powers in inverse horizon area (BHAF). Thus, we have assumed (a) that corrections to the BHAF are additive and (b) are primarily expressed as $log + inverse~ powers$ of the BHAF, with coefficients that depend on parameters involved in the quantization/modification of classical GR, to evaluate black hole entropy.

We now focus on the prediction of the very accurate observational validity of HAT following the BBHC event GW250114, to the corrections, via eqn.s (\ref{gslbb}) and (\ref{HATN}) 
\begin{eqnarray}
-\Delta s_{bh} & < &   (\Delta S_{BH})_{OBS} \nonumber \\
{\rm where}~ \Delta s_{bh} & \equiv & s_{bhr} - (s_{bh1} + s_{bh2})
\end{eqnarray}

\subsubsection{Absolute Consistency Criterion}

In view of the inequality (\ref{HATN}) we now propose the Absolute Consistency Criterion (ACC) \cite{pm2025}-\cite{pm2026}:
\begin{eqnarray}
\Delta s_{bh} > 0. \label{acc}
\end{eqnarray}
This is a {\it sufficiency} stipulation regarding the {\it algebraic sign} of the net change in corrections to the BHAF, rather than the magnitude, which we know, vide eqn (\ref{cor}), to be tiny in comparison with the BHAF. It is certainly not a {\it necessary} consequence of (\ref{HATN}), although, keeping eqn. (\ref{cor}) in view, it is clear that the ACC alone has the chance of non-trivial predictions; the alternative (i.e., negative) sign in the RHS of (\ref{acc}) compares two very disparate positive quantities. As per (\ref{cor}), this automatically guarantees consistency without consequences. On the other hand, the ACC has the potential for a richer range of ramifications on the parameters of the theory, as we explore in the sequel.  

The main properties of the ACC can now be summarised :

\begin{itemize}
\item {It is a sufficiency condition from BBHC-GW Observations.}
\item {Its algebraic sign is more important than its magnitude.}
\item {In this sense it is similar to HAT : $|\Delta A_h|$ unimportant}
\item {Reminiscent of Asymptotic Freedom in elementary particle theory : $ RG~~  \beta^{(1)}(g) < 0~,~ g < 1 $}
\item {In both these cases the algebraic sign is physically relevant, magnitudes are not}
\item {ACC may have important implications which ought to be explored. }
\item {The similarity between ACC and asymptotic freedom may not be coincidental; for Entanglement Entropy, the conformal (trace) anomaly may provide connection}
\end{itemize}

\section{Inverse Area Corrections to BHAF} 

\subsection{Wald-Jacobson-Kang-Myers Entropy Function}

The Lagrange density, in absence of matter fields. is denoted by $L=L(g_{ab}, R_{abcd})$, where $\nabla$ denotes appropriate spacetime covariant derivatives. A general variation of the classical action yields 
\begin{eqnarray}
\delta L &=& E\cdot \delta \phi + \nabla \cdot \Theta(\delta g_{ab}) \nonumber \\
\Rightarrow \nabla \cdot \Theta &=& \delta L - E_{ab} \delta g_{ab}~. \label{gen}
\end{eqnarray}         
Here, $E$ refers to the equation of motion terms resulting from the general variation of $\phi$, which vanish on-shell.

Consider now infinitesimal diffeomorphisms $x^a \rightarrow x^a + \xi^a(x)$; invariance of the classical action under such transformations imply that \cite{jkm1994}
\begin{eqnarray}
\Theta^a = 2 \frac{\partial L}{\partial R_{abcd}} \nabla_b(\nabla_d \xi_c + \nabla_c \xi_d) + \cdots  \label{diff}
\end{eqnarray}
where the dots imply terms with less than two derivatives on $\xi$. 
The corresponding N\"other current
\begin{eqnarray}
J_N^a &=& \Theta^a - \xi^a L \\ \label{nc}
\Rightarrow \nabla \cdot J_N &=& 0 ~{\rm when}~ E=0 . \label{con}
\end{eqnarray}
For diffeomorphisms, this implies that, once again ignoring less than second order derivative terms of $\xi$ (which is to be identified with the Killing vector field, for stationary isolated black holes), 
\begin{eqnarray}
J_N^a = \nabla_b \left(2 \frac{\partial L}{\partial R_{abcd}} (\nabla_d \xi_c) \right)
\end{eqnarray}
so that the N\"other charge may be written as 
\begin{eqnarray}
Q_N^{ab} = \left(2 \frac{\partial L}{\partial R_{abcd}} \nabla_d \xi_c \right) 
\end{eqnarray}
Identifying $\xi \rightarrow K$, and defining $l_{ab} \equiv \nabla_a K_b$ on the 2-spherical cross-section ${\cal S}$ of the horizon, the Wald entropy function is given by \cite{jkm1994}
\begin{eqnarray}
S_{bh} = \int_{\cal S} d^2a \frac{\partial L}{\partial R_{abcd}} l_{ab} l_{cd} \label{entr}
\end{eqnarray}   

\subsection{$F(R)$ Gravity}

\subsubsection{Area Law and corrections \cite{dmm2026}}

The motivation for classical modifications of general relativity has been a subject of much debate for decades, and there is no need here to revisit the arguments either in favour or against such modifications. Likewise, the rationale for an $F(R)$ replacement of the Einstein-Hilbert-Lorentz Lagrange density have also been discussed substantively in \cite{bre2004} - \cite{mod2011}. We realize from these important contributions that if a classically modified theory of gravity has black hole solutions like general relativity does, it is likely that the BHAF will receive modifications. For astrophysically macroscopic, spherically symmetric, stationary black hole solutions, with area of the horizon cross section $A_{\cal S} >> A_F$, where, $A_F$ is the analogue of the Planck area for $F(R)$ gravity, and deemed to be a fundamental constant of that theory, it is likely that \cite{pm2025}-\cite{pm2026}, the corrections to the area formula analogue of black hole entropy are additive
\begin{eqnarray}
S_{bh} = S_{BH} + s_{bh}(S_{BH}, Q) \label{cora}
\end{eqnarray} 
where, $Q$ represents collectively the parameters of both the theory as well as the black hole solutions under consideration. For astrophysical black holes, one may further stipulate that, since $S_{BH} >> 1$, the corrections admit an expansion in inverse powers of  the horizon area,
\begin{eqnarray}
s_{bh}(S_{BH},Q) = \sum_{n=1} s_n(Q) S_{BH}^{-n} ~. \label{expn} 
\end{eqnarray}  

We now demonstrate the explicit emergence of the BHAF and inverse area corrections for  macroscopic, spherically symmetric black holes of $F(R)$ gravity, using eqn. (\ref{entr}). It is clear that for such theories, we obtain
\begin{eqnarray}
S_{bh} = \int_{\cal S} d^2a F^{(1)}_R(R) l_{ab} l^{ab}~, \label{sfr}
\end{eqnarray} 
where $F_R^{(1)} = dF/dR$. Spherically symmetric static vacuum black holes of $F(R)$ gravity have been explored by \cite{seb2011}, and are characterized by the line element (in Schwarzschild coordinates)
\begin{eqnarray}
ds^2 = B(r) \exp 2\alpha(r) dt^2 - B^{-1}(r) dr^2 - r^2 d\Omega^2 \label{met} 
\end{eqnarray}
The Ricci scalar is given by \cite{seb2011}
\begin{eqnarray}
R(r) &=& 3 B_r \alpha_r + 2B \alpha_r^2 + B_{rr} + 2B \alpha_{rr} \nonumber \\
&+& 4r^{-1} \left[B_r + B \alpha_r \right] + 2r^{-2} (B-1)~. \label{ric}   
\end{eqnarray}
In eqn. (\ref{ric}), a subscript $r$ signifies a first order derivative of that function wrt $r$, while a double subscript signifies a second order derivative. The important point for us is to note that the Ricci scalar is an exclusive function of the radial coordinate $r$. It follows that $R(r_{\cal S}) = R_{\cal S}$, where $r_{\cal S}$ is the radius of the horizon, akin to the Schwarzschild radius for the Schwarzschild black hole in general relativity. Thus, eqn. (\ref{sfr}) can be rewritten as 
\begin{eqnarray}
S_{bh} = F^{(1)}_{R_{\cal S}}(R_{\cal S}) \int_{\cal S} d^2a~ l_{ab} l^{ab} =  F^{(1)}_{R_{\cal S}} (R_{\cal S}) A_{\cal S} ~\label{sfr2}
\end{eqnarray}
where $A_{\cal S}$ is the area of the horizon cross-section. Eqn. (\ref{sfr2}) coincides with eqn. (27) of the review article by Faraoni \cite{far2010}, based on the derivation and interpretations given in ref. \cite{bru2009}. Our eqn (\ref{sfr2}) of course specializes the more general $d$-dimensional result of \cite{bru2009} to four dimensional spherical black hole solutions of general $F(R)$ gravity, so that the inverse effective Newtonian coupling $G_{eff}^{-1}$ is identified with $F^{(1)}_{R_{\cal S}} (R_{\cal S})$. In our derivation, we have used the somewhat easier two codimensional horizon cross-section discussed in ref. \cite{jkm1994}, rather than the foliation of the black hole spacetime by bifurcation spheres of codimension 2, as per the original derivation in ref. \cite{rmw1992}-\cite{iw1994}. The advantage of our rederivation, even though appearing seventeen years following the original, is that it is readily applicable to comparison with astrophysical situations concerning GW data, corresponding to recent observations on BBHCs.    

To this end, observe that we can identify $R_{\cal S}$ in eqn. (\ref{sfr2}) with the {\it Gaussian} curvature of the two dimensional spherical cross-section of the horizon, so that it can be related to the cross-sectional area $A_{\cal S}$
\begin{eqnarray}
R_{\cal S} = \frac{a}{A_{\cal S}} \label{gau}
\end{eqnarray}  
where $a$ is a positive dimensionless number of $O(1)$. Henceforth, we shall absorb this number into a simple redefinition of $F(R)$, exactly as we have absorbed the factor 4 in eqn. (27) of ref. \cite{far2010}. 

To what extent is the identification of $R_{\cal S}$, i.e., the pullback of the Ricci scalar to ${\cal S} = h \cap \Sigma$ ($h$ is the null hypersurface associated with the black hole horizon and $\Sigma$ is a spacelike hypersurface foliating that horizon), with the Gaussian curvature of the two-sphere ${\cal S}$ justified ? To this end we note that there are no geometrical quantities other than this pullback to two-sphere ${\cal S}$, of the Ricci scalar of the spherical black hole solution, which has the right {\it dimensions}. If we think of a two dimensional formulation of $F(R)$ gravity on the foliation ${\cal S}$, then the identification (\ref{gau}) is certainly justified on such a two-sphere. So long as the full black hole geometry is a static spherically symmetric one, as being considered here, embedding such a two-sphere into the larger four dimensional spacetime may not bring in any extrinsic curvature effects which we may have missed. This argument is intuitive and subject to improvement upon more rigorous examination.        

In the limit of macroscopic black holes of very large area,  $R_{\cal S}$ is a very small quantity, so that $F_{R_{\cal S}}^{(1)}(R_{\cal S})$ can be Taylor expanded in powers of $R_{\cal S}$ around $R_{\cal S}=0$
\begin{eqnarray}
F^{(1)}_{R_{\cal S}}(R_{\cal S}) &=& \sum_{n=0}\frac{1}{n!} F^{(n+1)}_{R_{\cal S}}(0) R_{\cal S}^n \nonumber \\
&=& F^{(1)}_{R_{\cal S}}(0) + \sum_{n=1}\frac{1}{n!} F^{(n+1)}_{R_{\cal S}}(0) R_{\cal S}^n \label{tay}
\end{eqnarray}
where $  F^{(n)}_{R_{\cal S}}(0)~,~ n=1,2,...$ are the infinitely many parameters of the theory at vanishingly small values of the Ricci curvature.     

Substituting this expansion in eqn (\ref{sfr2}), we obtain
\begin{eqnarray}
S_{bh} = F_{R_{\cal S}}^{(1)}(0) A_{\cal S} + A_{\cal S} \sum_{n=1}\frac{1}{n!} F^{(n+1)}_{R_{\cal S}}(0) R_{\cal S}^n ~. \label{sfr3}
\end{eqnarray}
The first term can be made more akin to the BHAF provided $A_F^{-1} \equiv F^{(1)}_{R_{\cal S}}(0) > 0$; we define $S_{BH} \equiv A_{\cal S}/A_F$. Here, $A_F$ plays the role of the Planck area for $F(R)$ gravity, explicitly analogous to the effective Newtonian {\it constant} of $F(R)$ gravity, when such an identification can be made. This is expected to be small, on general intuitive grounds alone, so that $F^{(1)}_{R_{\cal S}}(0)$ is a large quantity of dimensions of inverse area. One can then define {\it corrections} to the BHAF by splitting off the first term in the expansion (\ref{sfr3}) as the BHAF, and identifying the rest of the expansion with the corrections,  
\begin{eqnarray}
S_{bh} &=& S_{BH} + s_{bh} \nonumber \\
s_{bh} &=& \sum_{n=1} s_{n-1} S_{BH}^{1-n} \nonumber \\
s_{n-1} & \equiv & \frac{1}{n!} [F_{R_{\cal S}}^{(1)}(0)]^{n-1}F_{R_{\cal S}}^{(n+1)}(0) \label{sfr4}
\end{eqnarray}  

\subsubsection{Implications}

Using eqn(\ref{sfr4}), and assuming that the coefficients $s_n$ are identical for similar mass black holes, the absolute consistency criterion implies
\begin{eqnarray}
&& \sum_{n=1}  s_{n-1} \left [ S_{BHr}^{1-n} - S_{BH1}^{1-n} - S_{BH2}^{1-n} \right] > 0 \nonumber \\
\Rightarrow && \sum_{n=1} s_{n-1} S_{BHr}^{1-n} \left[1 - (\frac{S_{BHr}}{S_{BH1}})^{n-1} - (\frac{S_{BHr}}{S_{BH2}})^{n-1} \right ] > 0 \label{inq} \\
\Rightarrow && \sum_{n=1}  s_{n-1} S_{BHr}^{1-n} [ 1 - \left( 1+ \frac{S_{BH1}}{S_{BH2}} \right) ^{n-1} \nonumber \\
&-& \left( 1+ \frac{S_{BH2}}{S_{BH1}} \right) ^{n-1}] > 0 \label{absc}
\end{eqnarray}
One may object to the above manipulations inside infinite series above, in view of the unknown sign of the coefficient $s_{n-1}$. However, the calculation is reliable for any {\it finite} truncation of the infinite series. In particular, we have for $n=2$, the result
\begin{eqnarray}
s_1 S_{BHr}^{-1} \left[ 1 - \left( 1 + \frac{S_{BH1}}{S_{BH2}} \right) - \left( 1 + \frac{S_{BH2}}{S_{BH1}} \right) \right]  >  0 \label{n2}
\end{eqnarray}      
Observe now that the quantity within the square brackets in the inequality (\ref{n2}) is actually $< -1$, so that one obtains from (\ref{absc}), using also (\ref{sfr4}), 
\begin{eqnarray}
S_{bh} = \frac{A_{\cal S}}{A_F} + F_R^{(2)}(0) + \frac12 F_R^{(3)}(0) A_{\cal S}^{-1} + \cdots, \label{abs2}   
\end{eqnarray}
which leads to, vide eqn (\ref{n2})
\begin{eqnarray}
s_1 = \frac12 F_R^{(3)}(0) F_R^{(1)}(0) < 0 \Rightarrow F_R^{(3)}(0) < 0 . \label{f3}
\end{eqnarray}
Thus if we restrict to the term $n=2$ in inequality (\ref{absc}), as may be valid for remnant black holes in BBHCs with very large horizon area, we obtain the result that $F^{(3)}_{R}(0) < 0$, i.e., for large enough areas of the remnant black hole, we obtain at least an algebraic sign constraint on this parameter of $F(R)$ gravity. This procedure can work for any finite order truncation $n$ of the inequality (\ref{abs2}), which, from a pragmatic standpoint is what may be required practically, depending on the magnitude of the remnant black hole horizon area. Thus, the absolute consistency criterion does indeed impose constraints on the parameters of the theory, as claimed earlier. Note that our focus here has been on the inverse area correction, and hence we have nothing to state about the arbitrary constant $F_R^{(2)}(0)$, given that the entropy functions are always defined upto an added constant. 

One might seek a clarification as to the limits and scope of our assay above, since the observational data on which (\ref{acc}) is based, deals with general relativistic analyses of {\it spinning} black hole coalescence. On the other hand, the anayses above involves non-spinning spherical black holes of $F(R)$ gravity, and not spinning generalizations thereof. Note however that the entire contentions above involve algebraic sign stipulations of very small, perturbative corrections to the BHAF calculated from the $F(R)$ gravity solution. Had such corrections been such that (\ref{cor}) is no longer valid, then perhaps the spin of the black hole would be an important factor for our consideration. Even so, it is not clear that the algebraic sign stipulations we have made on some $F(R)$ gravity parameters, would be invalidated under those circumstances.   

I now aditionally remark on the {\it scalaron} instability that has been discerned in the original literature on $F(R)$ gravity \cite{sot2010}, and recalled by an anonymous referee in their review. The scalaron field emerges upon a conformal transformation of the $F(R)$ action, transforming it to a {\it scalar-tensor} theory of gravity. It has been shown \cite{sot2010} that the theory can be rendered free of the ghost instability that plagues many higher derivative theories of gravity, by requiring that $F^{(2)}_{R}(0) > 0$. It is clear that HAT-validating GW data together with AC does not impose any restriction on $F^{(2)}_R(0)$ because the corresponding contribution to $S_{bh}$ is independent of $S_{BH}$ or $A_{\cal S}$. This term is therefore an arbitrary constant from our analyses, and in no way therefore contradicts the contentions of ref. \cite{sot2010} on the scalaron instability.

I should however remind the reader that contentions above were based solely on the spherical black hole solution (\ref{met}) for matter-free $F(R)$ gravity. I am not sure what the conformal transformation which maps the $F(R)$ theory to the scalar tensor theory, revealing the scalar degree of freedom, does to the spherical black hole solution (\ref{met}). So I cannot claim from the above analysis presented here that I have discerned any tangible reason for which $F(R)$ gravity is not a viable theory of gravity. Rather, the aim above has been to focus on the GW-observation inspired ACC and see what constraints it might impose on inverse area corrections to the BHAF, and hence indirectly on the parameters of the modified gravity theory. Proving viability of a theory or otherwise would entail a far more detailed investigation beyond the current scope of this review.       

\subsubsection{Subleading corrections to the BHAF : LQG modified It from Bit \cite{dmm2026}}

\begin{center}
\includegraphics[width=6cm]{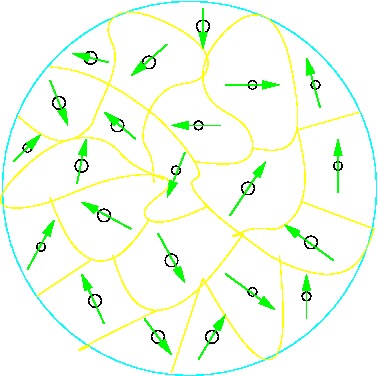}

\noindent {\small Fig. 5 Shows the spherical lattice covering the horizon with Planck-sized plaquettes populated by binary variables like spin$1/2$.}  
\end{center}

Since inverse area corrections to the BHAF for black hole entropy are not commonly a subject of discussion in the literature, we feel that we may mention another arena where such corrections make their appearance. This is in quantum general relativity. One of the more intuitive approaches towards a quantum general relativity rationale for the BHAF is the It from Bit approach formulated by Wheeler \cite{jaw1991}. In this approach, applicable to spherically symmetric black holes of general relativity, the horizon cross-section 2-sphere of area $A_{\cal S}$ is covered by a spherical lattice of plaquette size $A_{Pl}$, so that the number of plaquettes is $N \simeq A_{\cal S}/A_{Pl}$. On each of these plaquettes is placed a random binary variable, which can be spin $1/2$ variables for example. In this situation, the total number of spin $1/2$ states, i.e., the total number of microstates, on the horizon cross-section is given by 
\begin{eqnarray}
{\cal N} \simeq 2^N = 2^{A_{\cal S}/A_{Pl}}
\end{eqnarray}   

This leads to the Boltzmann entropy 
\begin{eqnarray}
S_{bh} = \log {\cal N} = \frac{A_{\cal S}}{A_{Pl}} ~\log 2 ~\label{jaw}
\end{eqnarray} 
The $\log 2$ coefficient is inevitable if the placing of random variables on the horizon uses binary variables. It can be absorbed into a redefinition of $A_{Pl}$, leading to the BHAF.

This entire scheme fits in nicely with the LQG counting of the microstates for non-rotating isolated black hole horizons \cite{dkm2001}, if {\it all} spins at punctures are assigned the value $1/2$. As is well-known, this contribution counts the number of ways of composing $N$ spin $1/2$ states. However, according to the basic framework of LQG, {\it not all} these states are physical, because they are not singlets under the symmetry group of spin networks namely $SU(2)$. The Gauss law constraint in time gauge (gauge-fixing away the Lorentz boosts) reduces to an $SU(2)$ constraint which only allows singlet states as physical states. Eqn.(\ref{jaw}) therefore represents an {\it overcounting} of the microstates  contributing to $S_{bh}$. 

To obtain the correct counting, consider a large, but even number of plaquettes $N$; one way to obtain a singlet is to count the number of microstates by placing spin $1/2$ components $\pm 1/2$ on half each of the total number of plaquettes. However, this too does not yield the precise number of singlet states, since microstates with total integral spin may also have such a distribution, and these must be excluded. The way to do this is to count the microstates by subtracting off from the $\pm 1/2$ distribution, the states with nonzero total integral spin, all of which have one extra spin $1/2$ component, either positive or negative. Such a counting leads to the formula
\begin{eqnarray}
{\cal N} &=& N! \left[ \frac{1}{[(N/2)!]^2} - \frac{1}{(N/2 +1)! (N/2 -1)!} \right] \nonumber \\
&=& \frac{N!}{(N/2 +1)! (N/2-1)! } \left( \frac{2}{N} \right)
\end{eqnarray}      

We now appeal to the Stirling approximation for the factorials for $N >> 1$,
\begin{eqnarray}
N! \simeq \sqrt{2\pi} \frac{N^{N+1/2}}{e^N} \label{sti}
\end{eqnarray}
This leads to the result
\begin{eqnarray}
{\cal N} & \simeq & \frac{1}{\sqrt{2\pi}} \frac{2^N}{N^{3/2}} \left( 1 + \frac{1}{2N} \right)^{-1} \nonumber \\
\Rightarrow S_{bh} &\simeq& S_{BH} - \frac{3}{2} \log S_{BH} - \frac12 S_{BH}^{-1} + \cdots \label{inva}
\end{eqnarray}
The important point in eqn. (\ref{inva}) is that the coefficient of the subleading inverse horizon area correction is {\it negative} thus in accord with our requirement of absolute consistency for quantum general relativistic black holes (see \cite{pm2024}-\cite{pm2025}). 

\section{Logarithmic Corrections to the BHAF \cite{pm2024}-\cite{pm2026}}

\subsection{Observational Preference}

We now turn to the logarithmic corrections to the BHAF which may ensue from {\it quantum} gravity (QG) proposals. Before turning to actual QG proposals, observe that the ACC vis-a-vis GW observational data validating the HAT, discussed in the previous sections leads to non-trivial constraints on the coefficients of logarithmic corrections. For astrophysical black holes, the ACC implies that $S_{BH} >> 1$ 
\begin{eqnarray}
\Delta s_{bh} &>& 0 \\ \nonumber
\Rightarrow \log \left[ \frac{ S_{BH1}^{s_{01}} S_{BH2}^{s_{02}}}{S_{BHr}^{s_{0r}}} \right] &<& 0 \label{logcor}
\end{eqnarray}
where, we have labelled the coefficient for the log correction according to the black hole it belongs. 

We now make the following assumptions/observation
\begin{itemize}
\item { $sig(s_{01,2,r})$, i.e., the algebraic sign of the coefficient is identical for all black holes}
\item { $|s_{01}| \sim |s_{02}| \sim |s_{0r}|$ i.e., the magnitudes are similar for all inspiralling black holes}
\item { $S_{BH1} \cdot S_{BH2} > S_{BHr}$} 
\end{itemize}
Clearly, the last of these is {\it not} an assumption, but a result borned out by BBHC data for GW150914, all the way upto GW250114. on the basis of these stipulations, we obtain
\begin{eqnarray}
sig(s_0) |s_0| \log \left[ \frac{ S_{BH1} S_{BH2} }{S_{BHr}} \right] < 0 \Rightarrow sig(s_0) = -
\end{eqnarray}
In other words, {\it GW Data validating HAT  chooses alg sign of $s_{bh}$ (QG) !}. This is what the ACC unequivocally leads to.  

\subsection{Theoretical contributions to $s_0$ : two proposals for QG}

We restrict our attention to two somewhat disparate proposals for QG : 
\begin{itemize}
\item {{\it Canonical QGR (LQG)} : non-perturbative background-free computation for quantum black hole with $A_h >> A_{Pl}$}
\item {{\it Entanglement Entropy (ENT)}: perturbative computation in classical black hole sptm background, based on Euclidean Quantum Gravity}
\end{itemize}
Ideally, we may want to combine the results of computations from both perspectives, so that a complete picture emerges. However, there are difficulties of combining the two QG proposals :
\begin{itemize}
\item {It is not possible to perform RG improvement of the Entanglement Entropy calculation (even after UV renormalization of Planck area) to match energy scales and renormalize  non-perturbative LQG-calculated Quantum black hole entropy, unlike Lattice-QCD renormalization through Perturbative-QCD.}
\item {Additional complication : LQG calculation is background-independent, while Entanglement entropy is calculated in classical black hole background.}  
\end{itemize}

\subsection{Loop Quantum Gravity \cite{t21982}-\cite{thie2007}}

\subsubsection{Basics}

LQG is a Canonical quantization attempt of GR which is background-independent and non-perturbative. The basic idea of this approach is to formulate it using the Hamiltonian formalism, quantizing the Arnowit-Deser-Misner approach to classical GR. However, rather than choosing the metric and affine connection as the dynamical variables, the canonical pair is often chosen to be the Levi-Civita connection, with the local boosts gauge-fixed, leaving a local $SU(2)$ rotation subgroup, and the corresponding densitized triad, i.e., the tetrad resticted to a spatial hypersurface. These two local variables obey the standard canonical commutation relation corresponding to $SU(2)$ gauge invariance, where these fields transform in the adjoint representation of the gauge group.

Now, because of the ultralocal nature of the canonical Heisenberg algebra, it is sometimes prefered to deal with $SU(2)$ group-valued {\it global} variables, respectively the holonomy and the flux, defined as 
\begin{eqnarray}
{\rm holonomy}~ h[{\bf A}]_C &=& \int_C {\bf A} \cdot {\bf dl}  \nonumber \\
{\rm fluxes}~ \Phi_S[{\cal F},f] &=& \int_S {\cal F}^I f_I~,~ {\cal F}^I = ({\bf E} \wedge {\bf E})^I~,I=1,2,3~.\label{gvar}
\end{eqnarray}
In eqn. (\ref{gvar}), ${\cal F}$ is the {\it solder 2-form} formed from the wedge product of the densitised triad one forms. $f$ is smooth function included to define the integral for the flux. This representation of spacetime depicts a {\it spin network} with edges $e$ which comprise the curve $C$, and vertices with $SU(2)$-invariant {\it intertwiners}. The flux of the densitised triad is over a spatial hypersurface $\Sigma$. We then have a visual characterization of `quantum space' which is a {\it closed} floating three dimensional lattice (like a three dimensional fishnet), with edges and vertices related to the holonomy (Wilson lines). Space exists only along the edges of this graph and at intertwiners. The `holes' in the lattice are supposed to be of Planck size, and are without geometrical significance. We color-code the visual characterization by the spin index of irreducible representations of the Little group of local rotations.
   
\begin{center}   
{\includegraphics[width=8cm]{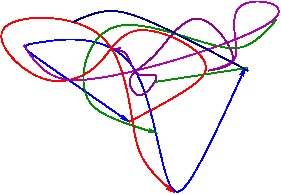} }

\noindent {\small Fig.5 A view of `Quantum Space' in terms of a typical spin network graph, whose edges define group-valued
Wilson lines and vertices are invariant $SU(2)$ tensors. The graph must be closed to reflext that it is a gauge singlet, and hence a physical state annihilated by the Gauss Law constraint operator. Likewise, the necessity of a `floating' rather than a rigid lattice is to ensure that the graph corresponds to a physical state annihilated by the Hamiltonian and momentum density operators, i.e., signifies spacetime {\it diffeomorphism} invariance. }
\end{center}

\subsubsection{Area Spectrum \cite{smo1989}-\cite{thie2007}}

The spin network basis of LQG is an {\it eigenbasis} for a self-adjoint {\it area operator}, with a {\it discrete} spectrum. To define the area operator, refer back to the floating lattice covering the two dimensional spherical horizon cross-section depicted in Fig.5. If each plaquette is defined to be a 2-surface $S_I,I=1,2, ...N$, with the area operator expressed as an integral over erach plaquette of the densitised triad projected down to the plaquette. The result is then summed over all plaquettes, leading to  
the discrete spectrum shown. 
\begin{eqnarray}
{\hat {A}}_S &\equiv & \sum_{I=1}^N \int_{S_I} {\det}^{1/2}[ ^2g({\hat E})] \label{aop} \\
a(j_1, \dots, j_N) &=& 8\pi \gamma l_P^2 \sum_{p=1}^N  \sqrt{j_p(j_p+1)} \label{aeig}
\end{eqnarray}
Now in the continuum limit, eespecially if the 2-surface is the cross-section of a (quantum) black hole, there are both lower and upper bounds on the area spectrum. 
\begin{eqnarray} 
\lim_{N \rightarrow \infty} a(j_1,....j_N) & \leq & {\cal A}_{cl} + O(l_P^2) ~{\rm for}~ j_p \leq \frac{k}{2} \\ 
& {\rm Equispaced} & \forall j_p=1/2
\end{eqnarray} 
\noindent{{\it Area Gap} : $a(\{j_I|I=1, ...,N\})$ has nonzero minimum value :
\begin{eqnarray}
a_{min}(j_1,j_2) = 8\pi \gamma l_P^2 \sqrt{3} 
\end{eqnarray}
The upper bound quoted above actually emerges from the $SU(2)$ Chern Simons (CS) gauge theory (coupled to spin-valued punctures of the bulk spin network, on the horizon) describing the horizon states of a quantum black hole in LQG. We describe this in slightly more detail in the next subsubsection. The Planck-scale area gap is a novelty in spacetime geometry. 

\subsubsection{Horizon degrees of freedom (DoF) and Dynamics} 
\begin{itemize}
\item In general a black hole horizon is a {\it null} hypersurface whic implies that the induced metric is {\it degenerate}.
\item Thus the dynamics on a black hole cannot involve standard quantum field theories where a non-degenerate metric is a necessity. Such dynamics must then be a {\it topological} quantum field theory. \item The horizon DoF for a non-rotating black hole are the restriction of the Levi-Civita connection (in time gauge with local Lorentz boosts gauge-fixed), i.e., the Bulk DoF, onto the horizon $\rightarrow SU(2)$ connection $A^I_a$; Solder 2-forms : ${\cal F}^I_{ab} \equiv \epsilon^{IJK} E_a^J E_b^K$ restricted (`pulled back') to the horizon.
\item Thus, the horizon DoF are basically $SU(2)$ Chern-Simons gauge fields, interacting with punctures from bulk spin network edges carrying spin $j_I~,~I=1, \dots,N$ as pointlike sources.
\item This enables us to write down the $SU(2)$ Chern Simons constraint equation as describing the horizon dynamics, in presence of spin-valued punctures as sources.
\begin{eqnarray}
\frac{k}{2\pi} F_{ab}^I = {\cal F}^I_{ab}~,~k \equiv \frac{A_{hor}}{A_P}~,~I = 1,2,3,...\label{cse}
\end{eqnarray}
\item The quantum description of such a horizon entails lifting eqn. (\ref{cse}) to the level of an operator equation acting on CS states carrying the spin of the relevant puncture. The Hilbert space of the horizon is thus the set of such CS states on the horizon cross-section with spin-valued punctures.
\end{itemize}

\subsubsection{Quantum Black Hole Entropy \cite{km98}-\cite{abhi-pm14}}

\begin{center}
\includegraphics[width=8cm]{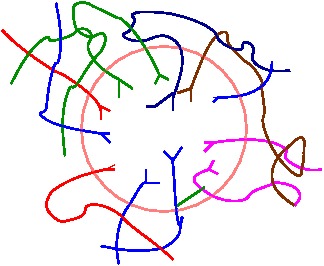}

\noindent {\small Fig.6 A non-rotating quantum isolated black hole}
\end{center}

To proceed to the issue of an {\it ab initio} counting of the horizon quantum states, i.e., the dimensionality of the Hilbert space of CS states with punctures, in order to obtain the black hole entropy, we recall its equivalence with the {\it number} of conformal blocks of a WZW model with punctures on the horizon cross-section :
\begin{eqnarray}
{\cal N}   & \equiv & dim {\cal H}_{CS + spins} \nonumber \\
\dim {\cal H}_{CS+spins} &=& \# [conf~ blcks]_{ [SU(2)~WZW]}~{\rm for}~k \equiv \frac{A_h}{A_{Pl}} >>> 1 \nonumber \\
S_{bh} & \equiv &   -Tr_{Blk} \rho \log \rho = \log {\cal N} . \label{sbh}
\end{eqnarray}
The important aspect of the final formula for $S_{bh}$ is that, it is the entropy of a {\it four} dimensional black hole, but the computation of this quantity is from the number of conformal blocks of a {\it two} dimensional conformal field theory, thus revealing the essential {\it holographic} content of this approach. Unlike other approaches, where the holographic hypothesis is {\it a priori} imposed to obtain black hole entropy, in this case, every step is rigorously derived from the preceding one. Thus, holography is evident in the formulation, and need not be imposed from the outside.  
For macroscopic black holes, the CS coupling constant $k \equiv A_h/A_{Pl} >> 1$; it is not difficult to show that, in this case
\begin{eqnarray}
{\cal N}(j_1, ...j_P)  &=& \prod_{i=1}^{P} \sum_{m_i=-j_i}^{j_i} [ \delta_{\sum_{n=1}^P m_n,0} \nonumber \\
&-& \frac12  \delta_{\sum_{n=1}^P m_n,-1} -\frac12 \delta_{\sum_{n=1}^P m_n,1} ] 
\end{eqnarray}
With $S_{bh} =\log {\cal N}$ we obtain
\begin{eqnarray}
S_{bh} = S_{BH} - \frac32 \log S_{BH} +{\cal O}(S_{BH}^{-1}) ~,~S_{BH} = \frac{A_{hor}}{A_{Pl}}
\end{eqnarray}
This therefore is an ab initio derivation of $S_{bh} \rightarrow$ LQG corrections to the BHAF. 

An alternative derivation of the same result, which does not make use of the details of CS theory Hilbert space, has already been provided here in section 4 : the modified It from Bit reported there. The logarithmic correction to the BHAF emerges there with an identical coefficient. Recall that, from the discussion on the obeisance of the result with the ACC earlier in this section, the obtained result certainly is in accord with the ACC with respect to GW observational data validating the HAT.

\subsubsection {Remarks on LQG log-area corrections to BHAF}
\begin{itemize}
\item $s_{bh} < 0 \Rightarrow $ black hole entropy calculations, whereby the corrected entropy is lower than the BHAF value, are absolutely consistent wrt GW observational data validating the HAT.
\item The result is exact for astrophysical black holes with $A_h/A_{Pl} >> 1$.
\item The corrections constitute a {\it finite} series without any ultraviolet renormalization ! 
\item We see that 4 dimensional black hole  entropy is computed from 2 dimensional CFT, which is the hallmark of holography, without having to invoke the so-called {\it holographic hypothesis} anywhere in the formulation.
\item Observe that this black hole entropy does not take interaction with matter into account. The inclusion of matter fluctuations is  hard, and may involve extra spinnet graph sets. 
\item The result is completely independent of the classical spacetime background. 
\item The one caveat in the calculation is the assumption that this kinematical result has been obtained assuming that quantum dynamical constraints preserve it; there is no complete proof of this property yet within LQG.  
\end{itemize}

\subsection{Entanglement Entropy \cite{cal1994}-\cite{cd1978}}

We begin with equilibrium statistical mechanics, invoking the canonical ensemble; ther partition function is defined as 
\begin{eqnarray}
Z(\beta) &=& tr \exp -\beta H \\
S(\beta) &=& -(\beta \partial_{\beta} -1) \log Z
\end{eqnarray}
For a spherically symmetric black hole, the spacetime factorizes as ${\cal M}_{1+1} \times S_{\perp}^2$. The time direction in the $1+1$ is next {\it Euclideanized} : $t \rightarrow i\tau$; in such a situation, a field theory partition function can be written over the fluctuating part of the metric ${\tilde g}$ and the matter field fluctuations, as 
\begin{eqnarray}
Z[g_{bh}] = \int {\cal D}{\tilde g} {\cal D} \Phi_m \exp - {\cal I}(g_{bh}, {\tilde g}, \Phi_m) 
\end{eqnarray}
\noindent For finite temp $\beta$, we require all fields in $Z$ to be periodic under $\tau \rightarrow \tau + \beta \Rightarrow Z[g_{bh}, \beta]$ develops conical singularities for arbitrary $\beta$ with deficit angle $\delta_{\beta} = 2\pi (1-\beta/\beta_H)$ at $h$, where $\beta_H^{-1} \rightarrow$ corresponds to the Hawking temperature (proportional to the surface gravity) for the background black hole spacetime metric $g_{bh}$ on $h$. The black hole entropy is defined then as \cite{cal1994}-\cite{sol2020} as 
\begin{eqnarray}
S_{bh}^{ent} = \left( 2\pi \frac{d}{d \delta_{\beta}} + 1 \right) \log Z(\delta_{\beta})|_{\delta_{\beta}=0} \label{enten}
\end{eqnarray}

\subsubsection{One loop contribution}

According to ref. \cite{sol2011}, the one loop calculation of $S^{ent}_{bh}$ is given {\it exactly} by the calculation of the conformal anomaly in a field theory of matter and graviton fluctuations in a classical background, by the formula derived by Christenson and Duff \cite{cd1978}, this leads to the following result for the logarithmic corrections beyond the BHAF \cite{sol2011}-\cite{sol2020}, after renormalization of $A_{Pl}$ (or $G$)
\begin{eqnarray}
s^{ent}_{bh} &=& s_0^{ent,1} \log S_{BH} + \cdots \nonumber \\
s_0^{ent,1} &=& \sum_{j=0}^{2} N_j a_j \label{1lp}
\end{eqnarray}
where, $N_j = \# ~{\rm species~of~spin}~ j=0^+,1/2,1,0^-,3/2,2~$. The coefficients $a_j$ are the oneloop contribution to the entanglement entropy of matter field fluctuations of spin $j$, which includes {\it graviton} fluctuations around the classical black hole background.

\subsubsection{Absolute Consistency}

Absolute consistency with HAT-validating GW Observational data imposes constraints on $N_j$: the one Loop Result for a four dimensional Schwarzschild spacetime background
\begin{eqnarray}
s^{ent,1}_0 &=& \frac{1}{45} \left[ N_{0^+} + \frac72 N_{1/2} - 13 N_1 \right] \nonumber  \\
&+& \frac{1}{45} \left[ 91 N_{0^-} -\frac{233}{4} N_{3/2} + 212 N_2 \right] \label{1lpsch}
\end{eqnarray}

\subsubsection{Combining LQG and Ent Entropy Log corrections; Caveats and Implications \cite{pm2026}}

Motivated by checking the ACC with GW data, we propose the {\it total} logarithmic correction to the BHAF, arising out of both the LQG and the entanglement entropy contributions, to be
\begin{eqnarray}
s_0 = s_0^{lqg} + s_0^{ent,1} \label{combo}
\end{eqnarray}
We recall that there are caveats in such an apparently blithe addition of the coefficients of the log corrections, however, as we shall comment later, there may be overlapping ideas between them. Nevertheless, from a physical standpoint, where both perturbative graviton and matter fluctuations, and non-perturbative quantum geometrical fluctuations are involved, given the smallness in the overall magnitude

\subsubsection{Implications \cite{pm2026}}

We now list the ramifications of the combination of the logarithmic correction coefficients given by eqn. (\ref{combo}), for various scenarios, including the SSET (see Introduction) of fundamental interactions, without and with graviton fluctuations, ignoring gravitini fluctations, but including some  Beyond-SSET (BSSET) species. 
\begin{itemize}
\item  SSET : $N_{0^+} = 1~,~ N_{1/2} =24~,~ N_1=12~,~N_{0^-} =0=N_{3/2}=N_2 \Rightarrow s_0 = -(277/90) < 0$! Thus, with no graviton graviton fluctuations, the SSET is absolutely consistent with the HAT-validating GW data. 
\item SSET + one BSSET Axion : $N_{0^-} = 1 \Rightarrow s_0 = -(95/90) < 0$ !
\end{itemize}
For these scenarios, results are in accord with ACC. However, consider now two BSSET scenarios
\begin{itemize} 
\item SSET + two BSSET Axions : $s_0 = +(29/30) > 0$
\item SSET + One axion + one graviton : $s_0 = 329/90 > 0$ 
\end{itemize}
Thus, in both these cases, there are issues of the calculated logarithmic corrections to the BHAF with the ACC.

It is obvious that if our stipulations made in the current subsection on logarithmic corrections to the BHAF for black hole entropy are correct, the demand of the ACC does appear to impose rather significant constraints on the BSET spectrum, including spin $2$ gravitons. As of now, there are no reported observational {\it results} (not bounds) on either gravitons or axions. 

\section{Summary and Discussion}

\begin{itemize}
\item Within a class of classical $F(R)$ gravity theories, namely those that are regular at vanishing small Ricci scalar curvature, demanding the ACC based on HAT-validating GW data, on the inverse horizon area corrections to the BHAF for the Wald entropy of spherical black holes, leads to a whole slew of relations between parameters of the modified gravity theory. 
\item Within quantum GR, combining results of LQG and ENT, we conclude that the SSET is absolutely consistent in our sense. A single additional spinless axion also appears to preserve this consistency. However, ACC appears to be at odds with a richer BSSET spectrum, including the existence of a spin $2$ graviton. Thus, if either a graviton, or a coexisting pair of axion species, of even an axion and a graviton, are observed in forthcoming astrophysical experiments, the ACC will stand to be falsified. This is interesting because of this direct implication that observations in the near future may have for our results, be they positive or negative. 
\item While we have somewhat na\"ively added on the coefficients of the logarithmic correction terms to the BHF, due to LQG and ENT, one may recall that there are methodological similarities in both approaches, namely both make use of {\it conical singularities} on the horizon to implement the counting to the black hole entropy corrections. The full implication of this fact has to be understood better. 
\item Gravitons are most likely very hard to observe at sub-Planckian energies, according to some analysis presented by F. Dyson \cite{dys2013}. Using ideas gleaned fropm some older work of G. 't Hooft \cite{thf1997}, a glimpse into the difficulty may be described as follows. From the dimensional Newtonian gravitational constant, one may derive two dimensionless energy-dependent coupling `parameters' : $g_s = Gs~,~ g_t = Gt$. How, at energies accessed typically in proton-proton collisions at the LHC, it is obvious that $g_s,g_t << 1$, implying that graviton cross section are expectedly negligible. For larger $g_s, g_t$ the cross-sections increase quadratically with the energy, so that for $g_s, g_t = {\cal O}(1) $ at Planckian energies. However, there may be issues with {\it perturbative unitarity} at such scales.  
\end{itemize}

\section{Acknowledgement} I thank the anonymous reviewer for raising several interesting points of clarification. I have tried to address these issues to the best of my ability.

\end{document}